\documentclass[aps,prc,twocolumn,superscriptaddress,amsfonts,amsmath,amssymb,showpacs,floatfix]{revtex4-1}
\usepackage[pdftex]{graphicx} 
\usepackage{hyperref,dcolumn,longtable}
\usepackage{color}
\usepackage[dvipsnames,svgnames,x11names,hyperref]{xcolor}
\hypersetup{colorlinks,citecolor=[RGB]{132 112 255},filecolor=[RGB]{220 20 60},linkcolor=[RGB]{191 62 255},urlcolor=[RGB]{138 43 226}}
\usepackage{float}
\usepackage{footnote}

\begin{document}


\title{Pairing properties of the double-{\boldmath$\beta$} emitter {\boldmath$^{116}$}Cd}
\author{D.~K.~Sharp}
\email[Correspondence to: ]{david.sharp@manchester.ac.uk}
 \affiliation{Schuster Laboratory, The University of Manchester, Oxford Road, Manchester, M13 9PL, UK}
  \author{S.~J.~Freeman}
 \affiliation{Schuster Laboratory, The University of Manchester, Oxford Road, Manchester, M13 9PL, UK}
  \author{B.~D.~Cropper}
 \affiliation{Schuster Laboratory, The University of Manchester, Oxford Road, Manchester, M13 9PL, UK}
  \author{P.~J.~Davies}
 \affiliation{Schuster Laboratory, The University of Manchester, Oxford Road, Manchester, M13 9PL, UK}
 \author{T.~Faestermann}
 \affiliation{Physik Department E12, Technische Universit\"at M\"unchen, D-85748 Garching, Germany}
  \affiliation{Maier-Leibnitz Laboratorium der M\"unchner Universit\"aten (MML), D-85748 Garching, Germany}
  \author{T.~M.~Hatfield}
 \affiliation{Schuster Laboratory, The University of Manchester, Oxford Road, Manchester, M13 9PL, UK}
  \author{R.~Hertenberger}
   \affiliation{Fakult\"at f\"ur Physik, Ludwig-Maximillians Universit\"at M\"unchen, D-85748 Garching, Germany}
  \author{S.~J.~F.~Hughes}
 \affiliation{Schuster Laboratory, The University of Manchester, Oxford Road, Manchester, M13 9PL, UK}
  \author{P.~T.~MacGregor}
 \affiliation{Schuster Laboratory, The University of Manchester, Oxford Road, Manchester, M13 9PL, UK}
   \author{H.-F~Wirth}
  \affiliation{Fakult\"at f\"ur Physik, Ludwig-Maximillians Universit\"at M\"unchen, D-85748 Garching, Germany}

\date{\today}

\begin{abstract}
The pairing properties of the neutrinoless double-$\beta$ decay candidate $^{116}$Cd have been investigated. Measurements of the two-neutron removal reactions on isotopes of $^{114,116}$Cd have been made in order to identify 0$^+$ strength in the residual nuclei up to $\approx$3~MeV. No significant $L=0$ strength has been found in excited states indicating that the Bardeen-Cooper-Schrieffer (BCS) approximation is a reasonable basis to describe the neutrons in the ground state. This approximation avoids complications in calculations of double-$\beta$ decay matrix elements that use the quasiparticle random-phase approximation (QRPA) techniques. However this is not the case for the protons, where pair vibrations are prevalent and the BCS approximation is no longer valid, complicating the use of traditional QRPA techniques for this system as a whole.
\end{abstract}

\maketitle


\section{\label{sec:level1}Introduction}
Neutrinoless double-$\beta$ decay ($0\nu2\beta$) is a second-order weak process that, if observed, indicates that the neutrino is a Majorana particle. A measurement of the decay rate would provide access to the absolute neutrino mass scale since, for a decay mitigated by a light Majorana neutrino,

\begin{equation}
(T^{0\nu}_{1/2})^{-1}=G_{0\nu}(Q_{\beta\beta},Z)|M_{0\nu}|^2\left<m_{\beta\beta}\right >^2,
\label{decay-rate}
\end{equation}

\noindent where $G_{0\nu}(Q_{\beta\beta},Z)$ is a phase-space factor, $M_{0\nu}$ is the nuclear matrix element (NME) describing the decay; and $\left<m_{\beta\beta}\right >$ is the effective Majorana mass of the neutrino~\cite{engel},

\begin{equation}
\left<m_{\beta\beta}\right >=\left| \sum_{k}m_k U^2_{ek}\right|.
\label{effective-mass}
\end{equation}

\noindent Here, $m_k$ are the neutrino mass eigenvalues and $U_{ek}$ is the electron row of the neutrino mixing matrix.

In order to extract a meaningful value for the neutrino mass from any future measurement of the decay rate, knowledge of the nuclear matrix element is required. These matrix elements are calculated using a range of theoretical frameworks which include the interacting shell model, interacting boson model and the quasiparticle random-phase approximation (QRPA), amongst others. Calculations using these different methods currently vary by factors of 2-3 in the calculated value of the NME for a particular $0\nu2\beta$ candidate (e.g., Ref.~\cite{engel}). As no other process samples the same matrix element, other experimental data are required to test the validity of the calculations.

QRPA calculations incorporate the pairing properties of nuclei by introducing like-particle pairing through the use of Bardeen-Cooper-Schrieffer (BCS) correlations. Indeed an analysis of this method has shown the importance of $J^{\pi}=0^+$ pairs to $0\nu2\beta$ decay~\cite{simkovic}. In this work, we will report on the pairing properties of the $A=116$ candidate system, $^{116}$Cd$\rightarrow^{116}$Sn. The two-neutron removal ($p$,$t$) reaction has been measured on a target of $^{116}$Cd with the aim of investigating the pair-transfer strength to excited $0^+$states. We also include data on $^{114}$Cd, used as a consistency check. In circumstances where the BCS approximation is valid, pair-transfer strength is dominated by the ground-state to ground-state transition with weak population of excited $0^+$ states. Strong population of $0^+$ states is indicative of shape transitions or pairing vibrations (Ref.~\cite{freeman} and references therein) which complicate the use of QRPA methods in calculating double-$\beta$ decay matrix elements. This work follows on from previous studies of the $A=76$~\cite{a=76n,a=76p}, $A=130$~\cite{a=130} and $A=100$~\cite{a=100} $0\nu2\beta$ candidate systems.

The double-$\beta$ decay with the emission of neutrinos ($2\nu2\beta$) of $^{116}$Cd has been studied most recently by the NEMO-3~\cite{nemo} and Aurora~\cite{aurora} experiments. The latter used enriched $^{116}$CdWO$_4$ crystal scintillators and has the most precise measurement of the ground-state to ground-state $2\nu2\beta$ half-life [$T_{1/2}=(2.63^{+0.11}_{-0.12})\times10^{19}$~yr] and set an improved limit on the $0\nu2\beta$ decay of $T_{1/2} \geq 2.2\times10^{23}$~yr. $^{116}$Cd is also proposed as a candidate as part of a future multi-isotope bolometric experiment, the CUORE Upgrade with Particle IDentification, to search for $0\nu2\beta$~\cite{cupid}.

The ($p$,$t$) reactions have been measured previously on the cadmium isotopes. However, none of these studies highlighted the distribution of $0^+$ states, instead focusing on either ground-state transitions~\cite{pt-bassani,pt-bauer}, the first $2^+$ states~\cite{pt-comfort} or on deep-lying orbits~\cite{pt-crawley}. As such there is little information on excited $0^+$ states following these reactions on either $^{114}$Cd  or $^{116}$Cd. Data exist on pair-correlated states in the residual nuclei via measurements of the ($t$,$p$) reactions on targets of $^{112}$Cd~\cite{112cd-tp} and $^{114}$Cd~\cite{114cd-tp}, which are discussed below. In the case of the residual $^{112}$Cd nucleus, high-resolution ($p$,$p'$) and ($d$,$d'$) data exist that provide information on spin assignments for excited states in this nucleus, including the location of excited $0^+$ states~\cite{hertenberger}. The current work reports on states populated in the $^{114,116}$Cd($p$,$t$)$^{112,114}$Cd reactions up to $\approx$3~MeV in excitation and with superior resolution compared with the limited data previously available in the literature.

\section{\label{sec:level1}Experimental Methods}
 
These measurements were made at the Maier-Leibnitz Laboratorium (MLL) of the Ludwig Maximilians Universit\"at and the Technische Universit\"at M\"unchen, where the MP tandem accelerator provided a beam of protons at an energy of 22~MeV with a current of $\approx$1~$\mu$A. This beam was used to bombard isotopicaly-enriched metallic targets of $^{114}$Cd (98.55\%) and $^{116}$Cd (98.07\%), with a nominal thickness of $\approx$50~$\mu$g/cm$^2$, mounted on a carbon backing $\approx$10~$\mu$g/cm$^2$. The beam dose was monitored throughout the experiment using a Faraday cup connected to a current integrator. 

Ejectile ions from reactions on these targets were momentum analyzed using the Q3D magnetic spectrograph~\cite{Q3D}.  The ions were identified at the focal plane of the spectrograph using a combination of two proportional counters backed by a plastic scintillator~\cite{Wirth} which provided signals proportional to energy-loss and residual energy deposited. The position of the ion trajectories along the focal plane was determined by reading out 255 cathode pads equipped with individual integrated preamplifiers and shapers, with 3.5mm pitch, positioned along the length of one of the proportional counters. An event was triggered if three to seven adjacent pads registered a signal above threshold. Position is then determined by fitting a Gaussian line shape to the digitized signals from the active pads. This method gave a position resolution of better than 0.1~mm. Ejectiles from the reaction of interest were identified though comparison of their energy loss in the proportional counters and plastic scintillator as well as their focal-plane position. The entrance aperture of the spectrograph was fixed at values of either 14.03~msr (full aperture) or 7.25~msr (half aperture) during the experiment.

In order to extract absolute cross sections, measurements of the product of the target thickness and spectrometer entrance aperture for each target and each aperture setting were made using Coulomb elastic scattering of 9-MeV deuterons at $\theta_{lab}$=20$^\circ$. Under these conditions, the cross section is within 2\% of Rutherford scattering, as calculated using the deuteron optical potential from Ref.~\cite{deuteron}. The beam currents for these measurements were much less than in the ($p$,$t$) reaction measurements; the different integrator scales were calibrated using a constant current source.

\begin{figure}[h]
\centering
\includegraphics[width=\columnwidth]{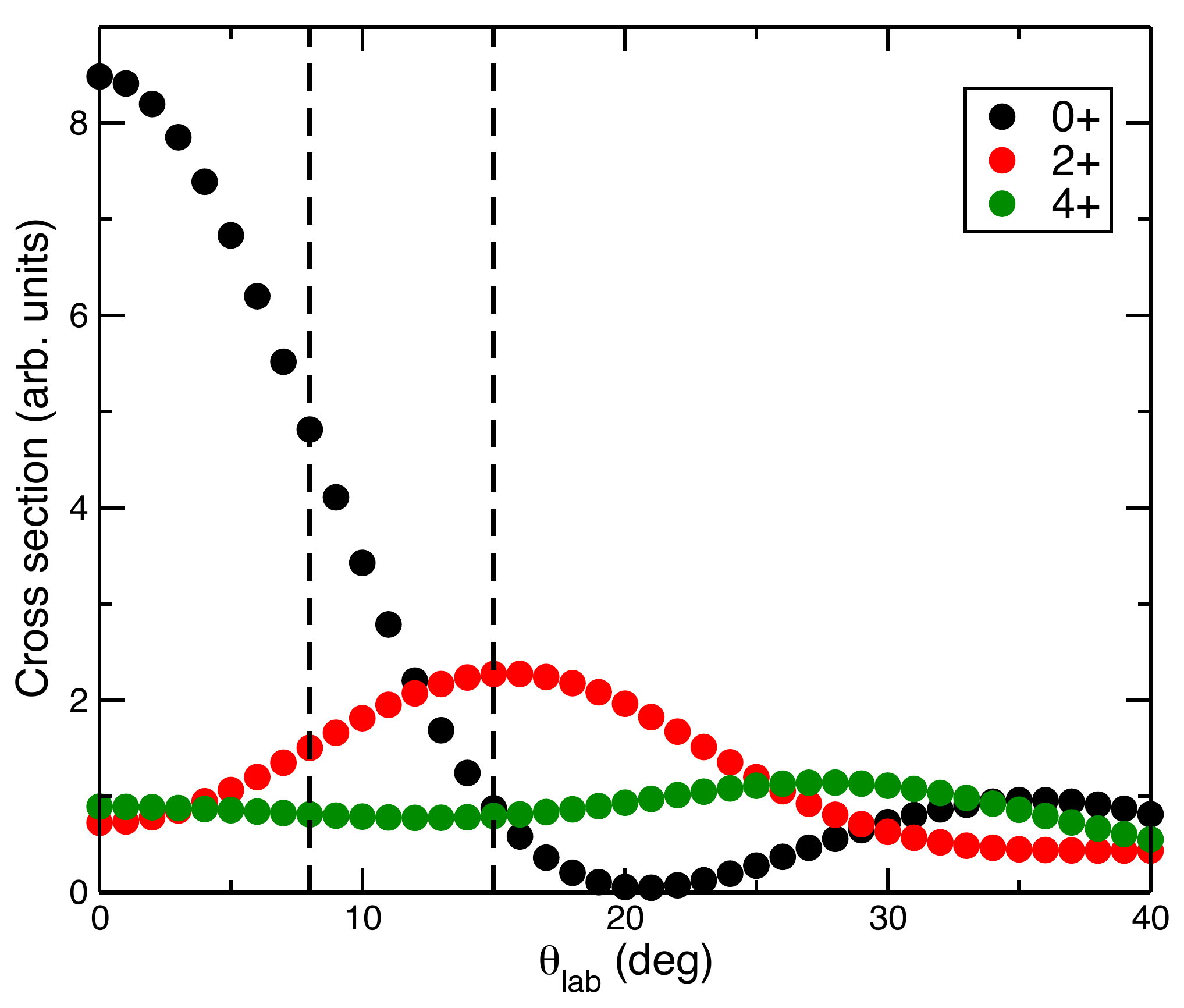}
\caption{\label{DWBA} Calculated angular distributions for the $^{116}$Cd($p$,$t$)$^{114}$Cd reaction for $L=0$ (black), 2 (red) and 4 (green) transfer to $0^+, 2^+$ and $4^+$ states in the residual nucleus. Dashed lines highlight the angles where cross sections were measured in this work. Details on the calculations can be found in the main body of the text.}
\end{figure}

Figure~\ref{DWBA} shows calculated distributions for the $^{116}$Cd($p$,$t$)$^{114}$Cd reaction calculated using the code and parameters discussed below. Data were taken at laboratory angles of $8^{\circ}$ and $15^{\circ}$. Ideally a measurement at $0^{\circ}$ would be preferred but due to count rate limitations this was not possible. The second angle of $15^{\circ}$ was chosen at the peak for an $L=2$ transfer. The angular distributions illustrate that the ratio of the cross sections at these two angles can discriminate the $L=0$ transfer from higher $L$ transfer, which are not peaked at forward angles. The ratio of cross sections at  $8^{\circ}$/$15^{\circ}$ would be $>1$ for $L=0$ and $\lesssim 1$ for $L>0$. 

\section{\label{sec:level1}Pair-transfer reactions}
\begin{figure}[h]
\centering
\includegraphics[width=\columnwidth]{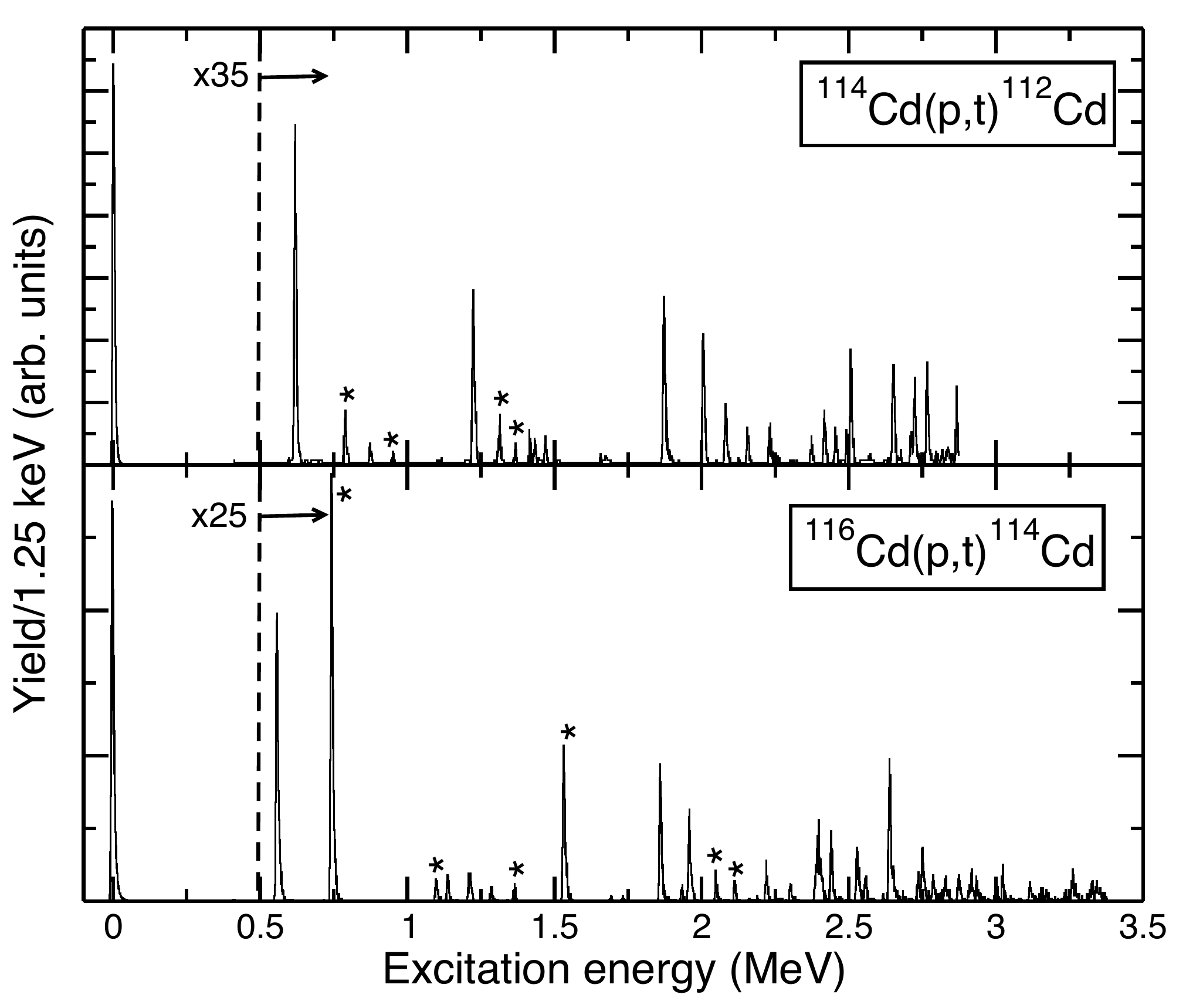}
\caption{\label{ex} Excitation energy spectra for the residual nuclei populated via the ($p$,$t$) reactions on $^{114}$Cd and $^{116}$Cd targets. Sections of the spectra to the right of the dashed vertical lines have been multiplied by the indicated factor. Peaks identified from reactions on contaminants in the target are labeled with an asterisk (*). }
\end{figure}
States in the residual nuclei following the removal of two neutrons were measured up to an excitation energy of $\approx$3~MeV, using five magnetic field settings of the spectrograph. The excitation spectra for states populated in $^{112,114}$Cd are shown in Figure~\ref{ex}, where the energy resolution obtained was $\approx$8~keV full width at half maximum. The excitation energies of the populated states have been calibrated using known states~\cite{nndc-114,nndc-112} and in most cases are known to better than 5~keV. States from reactions on isotopic contaminants in the target were identified using measured magnetic rigidities.

\begin{table}[h]
\caption{\label{0+} Summary of $0^+$ states identified in $^{112}$Cd and $^{114}$Cd via the ($p$,$t$) reactions measured in the current work. $0^+$ states identified in $^{116}$Sn and $^{118}$Sn via the ($^3$He,$n$) reaction measurements of Ref.~\cite{hn} on the same targets are also summarized. The cross sections measured at a forward angle are given along with the relative intensity compared to the ground-state transition from $^{114}$Cd, normalised for the $Q$-value dependence of the reaction cross section as described in the text. States marked with an asterisk are new $0^+$ states not previously reported in the literature or without a firm $0^+$ assignment.\\}
\begin{ruledtabular}
\begin{tabular}{ l l l l}

 &Ex (MeV)&$\sigma _{8^{\circ}}$($mb/sr$)&$I_{rel}$ (\%)\\
\hline
\rule{0pt}{3ex}  $^{114}$Cd($p$,$t$)$^{112}$Cd&0.00	& 4.226(20)& 	100\\
&1.225(1)	& 0.060(2)&		1.68\\
&1.432(1)	& 0.006(1)&		0.16\\
&1.872(1)	& 0.066(2)&		2.15\\
&2.648(1)	& 0.035(2)&		1.44\\
\hline 

\rule{0pt}{3ex}  $^{116}$Cd($p$,$t$)$^{114}$Cd&0.00	& 4.028(19)& 	98.06\\
&1.135(1)	& 0.014(1)&		0.38\\
&1.860(2)	& 0.068(2)&		2.07\\
&2.438(2)	& 0.035(1)&		1.19\\
&2.547(4)	& 0.015(1)&		0.53\\
&2.638(1)	& 0.079(2)&		2.88\\
&2.832(2)$^*$	& 0.015(1)&		0.58\\
&3.253(3)$^*$	& 0.026(2)&		1.17\\ 
\hline \hline
 &Ex (MeV)&$\sigma _{0^{\circ}}$($mb/sr$)&$I_{rel}$ (\%)\\
\hline
\rule{0pt}{3ex}  $^{114}$Cd($^3$He,$n$)$^{116}$Sn&0.00	& 0.139& 	100\\
&1.840	& 0.138&		71.91\\
&3.420	& 0.037&		14.93\\
&4.320	& 0.070&		24.98\\
&4.940	& 0.051&		17.62\\
\hline 
\rule{0pt}{3ex}  $^{116}$Cd($^3$He,$n$)$^{118}$Sn&0.00	& 0.091& 	83.07\\
&1.770	& 0.115&		76.14\\
&3.020& 0.016&		8.55\\
&4.450	& 0.055&		23.36\\

 \end{tabular}
 \end{ruledtabular}
 \end{table}
 
\begin{figure}[h]
\centering
\includegraphics[width=\columnwidth]{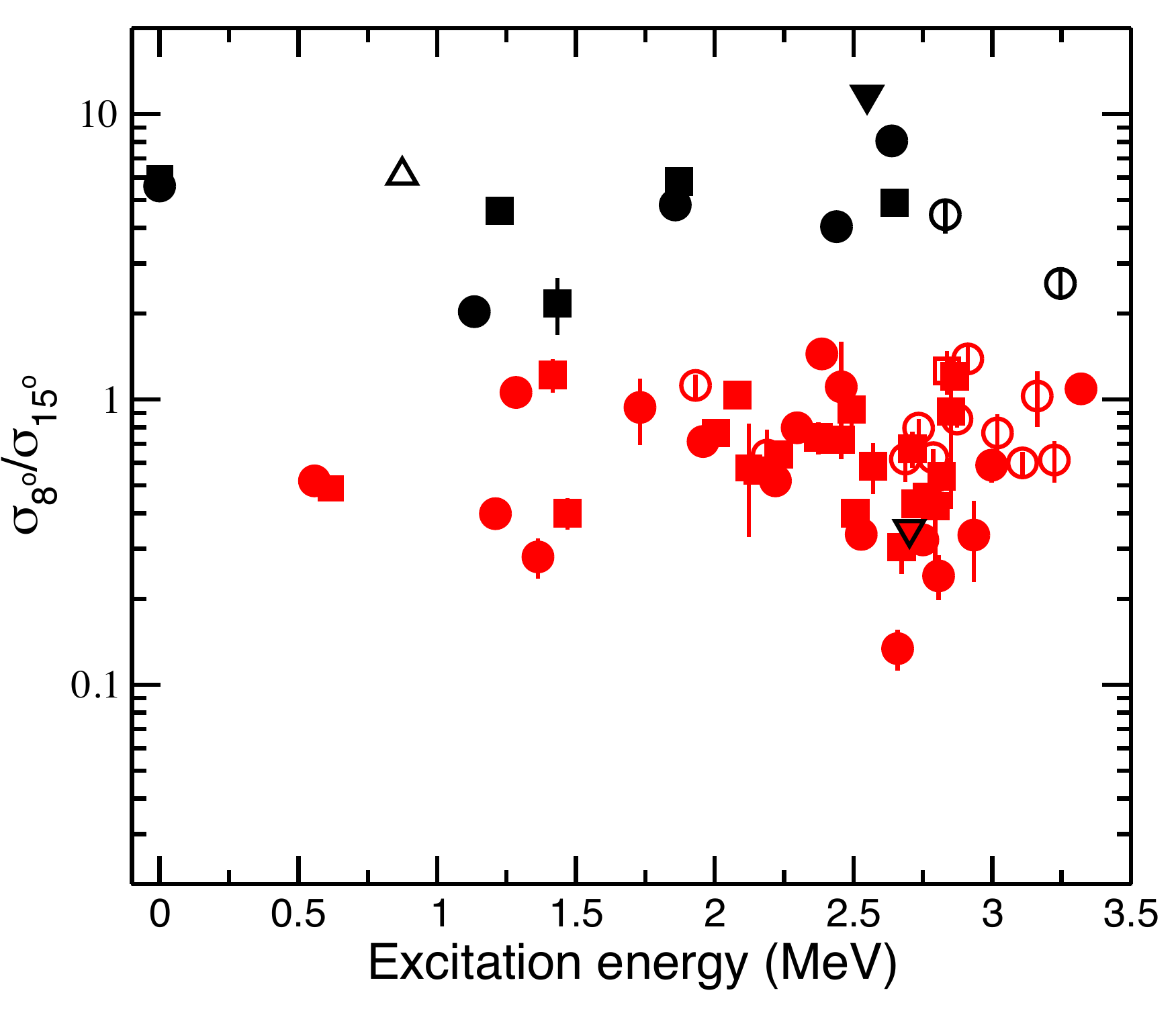}
\caption{\label{ratio} Ratio of cross sections measured at 8$^{\circ}$ and 15$^{\circ}$ for reactions on $^{114}$Cd (squares) and $^{116}$Cd (circles). Solid black points are known $L=0$ states in the residual nuclei, while solid red points are known states with $L > 0$. Empty points are states with no previous firm assignment in the literature. States in $^{112}$Cd (up-triangle) and $^{114}$Cd (down-triangle) where cross sections at 15$^{\circ}$ could only be assigned as an upper limit, and therefore the ratio is a minimum limit, are included.}
\end{figure}

The ratio of the cross section at the two measured angles was taken and is shown in Figure~\ref{ratio}. This ratio was greater than 2.0 for all previously known $0^+$ states. This criterion was used to assign new $0^+$ states, which are given in Table~\ref{0+}.  There were three states, one populated in $^{112}$Cd (0.874~MeV) and two in $^{114}$Cd (2.547 and 2.703~MeV), where only a limit on the cross section could be extracted at 15$^{\circ}$ due to low yield. For these states, only a lower limit can be placed on the cross-section ratio. However, this lower limit still exceeds that expected for $L=0$ for the 0.874~MeV state in $^{112}$Cd and the 2.547~MeV state in $^{114}$Cd.  

A total of two new $0^+$ states have been identified in $^{114}$Cd. A further state with a large angle ratio was observed at 0.874~MeV in $^{112}$Cd. This has not been observed in high-resolution ($p$,$p'$) and ($d$,$d'$) measurements of Ref.~\cite{hertenberger}, the ($d$,$p$) reaction of Ref.~\cite{dp} nor via nonselective reactions such as ($n$,$\gamma$)~\cite{ngam} and ($n$,$n'$$\gamma$)~\cite{garrett}. It would be very surprising to identify a new state at such low excitation energy in a well-studied nucleus and so, whilst there is no obvious contaminant identified to account for it, given the wealth of previous data this state has not been assigned as a $0^+$ state in $^{112}$Cd. 

The intensities of the states have been corrected for the $Q$-value dependence on the reaction cross section and expressed as an intensity relative to the ground-state transition in the reaction on $^{114}$Cd. This has been done by calculating the reaction cross-section as a function of excitation energy in the distorted-wave Born-aproximation framework using the code Ptolemy~\cite{ptolemy}. Global optical-model parameters for the proton and triton were taken from Refs.~\cite{koning} and \cite{pang}, respectively. The $L=0$ neutron pair is bound, either to the core of the target or the proton, with an energy equal to the corresponding two-neutron separation energy. The configuration of the neutron pair was chosen such that the bound state form factor has the appropriate number of nodes for pair removal from the $sdg$ shell only. Whilst other choices might be made in the reaction modeling, the calculations performed here are aimed at only removing $Q$-value dependence from the measured cross section to infer relative intensities. In order to assess the effect of the choice of nodes for the dineutron, the calculations were also performed for removal from the $h_{11/2}$ orbital, which can be occupied in the mid-shell cadmium isotopes. This gave a reduction in the relative intensities, due to the difference in $Q$-value dependence, shown in Table~\ref{0+} of up to 20\%, but does not significantly change the conclusions drawn below.

It is clear from the relative intensities in Table~\ref{0+} that there is no population of excited $0^+$ states with a strength of more than 3\% of the ground-state population. The cross section for the population of the ground state on each target are the same to within 5\%. A full compilation of the deduced energies and measured absolute cross sections for all states from reactions on both targets is given in the Supplementary Material~\cite{supp}.  

An equivalent analysis of the published cross-sections from previous ($t$,$p$) reaction studies~\cite{112cd-tp,114cd-tp} was performed. Excited states in the same residual nuclei are populated by two-neutron addition with no more than $\approx$6\% of the ground-state population. The lack of significant strength in excited $0^+$ states in any of these reactions indicates an absence of pair vibrational effects and helps to establish the validity of the BCS approximation for neutrons in these cadmium isotopes.

A $0\nu2\beta$ decay of $^{116}$Cd would lead to a $^{116}$Sn daughter nucleus. Neutron pairing in tin nuclei follows a classic BCS picture. A previous systematic study of the ($p$,$t$) reactions on the stable even-mass tin isotopes exists~\cite{pt-tin}. The experimental conditions were similar to the current work with an incident beam energy of 20~MeV, probing up to an excitation energy of $\approx$3~MeV and with $Q$-value resolution of $\approx$25~keV. Very little strength in excited $0^+$ states was observed ($<3\%$ of the ground-state transition), consistent with the superfluid nature of neutrons in the even tin isotopes. A BCS approach appears valid for neutrons in both parent and daughter nuclei in calculations of the $0\nu2\beta$ matrix element for decays of $^{116}$Cd.

While these data demonstrate that the neutrons in the $A=116$ $0\nu2\beta$ system follow the classic BCS picture, the same cannot be said for the protons. Previous proton pair-transfer data exist on the cadmium isotopes via a measurement of the ($^3$He,$n$) reactions~\cite{hn} and
the methods used above have been applied to the published cross sections. Table~\ref{0+} summarizes this data giving intensities relative to the ground-state transition from measurement of the ($^3$He,$n$) reaction on $^{114}$Cd. Here the global optical-model parameters for $^3$He and the neutron were taken from Refs.~\cite{pang} and \cite{koning}. The nodes were again chosen for pair addition to the $sdg$ shell. There is a much greater difference in the absolute cross sections for the ground-state transitions for two-proton addition on $^{114}$Cd and $^{116}$Cd, of $\approx$35\%, and the transitions to the first excited $0^+$ state in each have cross sections essentially equal to or greater than that of the ground-state transitions. The relative intensities show significant fragmentation of the $0^+$ strength over a number of states. These strong excited transitions are fragments of the pairing vibrational state~\cite{hn}. The centroids of these fragments appear at lower energies than expected from simple pair-vibrational models and indicate the need to consider particle-hole interactions in calculations as discussed in detail in Ref.~\cite{flynn}. These phenomena are not consistent with the assumptions of QRPA theory. 

The situation in the cadmium isotopes is similar to the pairing properties of the $^{128,130}$Te $0\nu2\beta$-candidate systems, where proton-pairing vibrations occur due to the $Z=64$ sub-shell closure~\cite{hn-te} but where the neutrons follow the BCS description~\cite{a=130}. If the BCS treatment near closed shells is unreliable then adaptations to the QRPA approaches are needed. For example, in the Te isotopes, the work of Ref.~\cite{bes} uses a hybrid model. The superfluid picture for neutrons remains but the protons are described as one- or two-pairing-phonon states treated in a normal phase and in the isovector vibrational model. These calculations reproduce the experimental NME for the $2\nu2\beta$-decays of $^{128,130}$Te. Similar approaches could be applied to the $A=116$ system. Indeed it should be noted that similar problems could also occur in the treatment of the decays of $^{124}$Sn and $^{136}$Xe~\cite{freeman}.

\vspace{-0.5cm}
\section{\label{sec:level1}Summary}
In summary, the distribution of pair-transfer strength has been measured for two-neutron removal from the $0\nu2\beta$-decay candidate $^{116}$Cd, along with $^{114}$Cd as a consistency check. Excited $0^+$ states are populated at the level of less than 3\% of the ground-state $L=0$ transition. This indicates that the neutrons, in isolation, in these cadmium isotopes exhibit a classic superfluid nature such that the BCS description of neutrons for these nuclei remains valid. However, the situation for protons is very different with pairing vibrations prevalent in these systems. This system, like others near to proton shell-closures, is likely to pose a problem for traditional QRPA calculations of the NME for $0\nu2\beta$ decay. Benchmarking these types of calculations, as well as those from other methods, against other nuclear data, such as the neutron and proton occupancies or the observed pair-transfer strength, could provide further information in order to assess their robustness.

\section{ACKNOWLEDGEMENTS}
We thank the staff at the Maier-Leibnitz Laboratorium der M\"unchner Universit\"aten for their assistance in accelerator operation and target manufacture. This material is based upon work supported by the UK Science and Technology Facilities Council and the Deutsche Forschungsgemeinschaft Cluster of Excellence ``Origin and Structure of the Universe".


\end{document}